\documentclass[preprint,showpacs,aps]{revtex4} 
\input{epsf}  
\usepackage{graphicx}
\usepackage{epic}
\usepackage{graphpap}
\usepackage{epic}
\usepackage{eepic}
\usepackage{color}
\usepackage{amssymb}
\usepackage{amsmath}
 
\begin{document}  

\title{Binomial moment equations for stochastic reaction systems} 
\author{Baruch Barzel and Ofer Biham}  

\affiliation{  
Racah Institute of Physics,   
The Hebrew University,   
Jerusalem 91904,   
Israel}  

\renewcommand{\P}[2]
{
P_{\vec #1}^{\vec #2}
}

\newcommand{\arrow}[1]
{
\overset{#1}{\longrightarrow}
}

\newcommand{\av}[1]
{
\langle #1 \rangle
}

\newcommand{\W}[1]
{
\langle
{\bf W}_{\vec #1}
\rangle
}

\newcommand{\Wm}[1]
{
\langle
{\bf W}_{#1}
\rangle
}

\newcommand{\WN}[1]
{
{\bf W}({\vec N},{\vec #1})
}

\newcommand{\A}[2]
{
{\bf A}_{#1 #2}
}

\newcommand{\Z}[2]
{
{\bf Z}_{#1 #2}
}

\newcommand{\z}[1]
{
{\bf Z}_{#1}
}

\newcommand{\B}[2]
{
{B}_{\vec #1}^{\vec #2}
}

\newcommand{\T}[1]
{
{T}_{\vec #1}
}

\newcommand{\F}[2]
{
{\bf F}^{\vec #2}_{#1}
}

\newcommand{\N}[1]
{
\langle N_{#1} \rangle
}

\newcommand{\Nt}[1]
{
\langle N_{#1} \rangle_t
}

\newcommand{\Ns}[1]
{
\langle N_{#1}^2 \rangle
}

\newcommand{\Nc}[1]
{
\langle N_{#1}^3 \rangle
}

\newcommand{\NN}[2]
{
\langle N_{#1}N_{#2} \rangle
}

\newcommand{\NNN}[3]
{
\langle N_{#1}N_{#2}N_{#3} \rangle
}

\newcommand{\NiNj}[2]
{
\langle N_i^{#1} N_j^{#2} \rangle
}

\newcommand{\NisNj}
{
\langle N_i^2 N_j \rangle
}

\newcommand{\NiNjs}
{
\langle N_i N_j^2 \rangle
}

\newcommand{\NisNk}
{
\langle N_i^2 N_k \rangle
}

\newcommand{\NjsNk}
{
\langle N_j^2 N_k \rangle
}

\newcommand{\NiNks}
{
\langle N_i N_k^2 \rangle
}

\newcommand{\NjNks}
{
\langle N_j N_k^2 \rangle
}

\newcommand{\R}[2]
{
\langle R_{#1 #2} \rangle
}

\newcommand{\Q}[1]
{
{\mathcal Q}_{\vec #1} 
}

\newcommand{\Qm}[1]
{
{\mathcal Q}_{#1} 
}

\newcommand{\D}[2]
{
{\vec{\bf D}}_{#1}^{#2}
}

\newcommand{\G}[3]
{
{\vec{\bf G}}_{#1 #2}^{#3}
}

\newcommand{\PG}[3]
{
{\bf P}_{#1 #2}^{#3}
}

\newcommand{\PD}[2]
{
{\bf P}_{#1}^{#2}
}

\newcommand{\K}[3]
{
{\bf K}_{#1 #2}^{#3}
}

\newcommand{\M}[4]
{
{\bf m}_{#1 #2 #3}^{#4}
}

\newcommand{\m}[3]
{
{\bf m}_{#1 #2}^{#3}
}

\newcommand{\Bm}[4]
{
{\bf B}_{#1 #2}^{#3 #4}
}

\newcommand{\Fm}[3]
{
{\bf F}_{#1}^{#2 #3}
}

\newcommand{\Prob}
{
P(\vec N)
}

\newcommand{\Inc}[1]
{
{\mathcal S}^{#1}
}

\newcommand{\I}
{
{\mathcal I}
}

\newcommand{\Pairs}[2]
{
{\mathcal N_{#1 #2}}
}

\newcommand{\Singles}[1]
{
{\mathcal N_{#1}}
}

\begin{abstract}

A highly efficient formulation of moment equations 
for stochastic reaction networks is introduced.
It is based on a set of binomial moments that
capture the combinatorics of the reaction processes.
The resulting set of equations can be easily truncated
to include moments up to any desired order. 
The number of equations
is dramatically reduced compared to the master equation.
This formulation enables the simulation of complex reaction
networks, involving a large number of reactive species much
beyond the feasibility limit of any existing method.
It provides an equation-based paradigm to the analysis of
stochastic networks, complementing the commonly used
Monte Carlo simulations.

\end{abstract}


\maketitle

Stochastic reaction networks appear in many natural systems,  
from grain-surface chemistry in the interstellar medium
\cite{Charnley1998}
to biochemistry in cells 
\cite{McAdams1997,Kaern2005}
and to ecological systems
\cite{Lande2003}.
In order to characterize these networks one wishes to evaluate
the time dependent 
populations of the reactive species. 
Due to the large fluctuations in stochastic systems,
the rate equations fail and stochastic methods 
\cite{vanKampen1981,Gardiner1985}
are required.
These methods are based on the master equation,
which can be solved either by direct integration
or
by Monte Carlo simulations
\cite{Gillespie1977,Gillespie2007}.
The problem with these methods is that they quickly become infeasible
as the number of reactive species increases.
More specifically, the number of equations in the master equation
proliferates exponentially with the number of species, 
making the stochastic analysis 
difficult even for empirical networks of moderate size
\cite{Stantcheva2003,Lipshtat2004}.
A common approach is to derive moment equations 
by tracing over the master equation
\cite{McQuarrie1967}.
In these equations one directly computes different moments related 
to the populations of the reactive species.
The problem is that higher order moments appear 
on the right hand side of the 
equations, making it difficult to obtain a closed set of equations.
In most closure schemes,
only the first and second moments are included
\cite{Gomez-Uribe2007}.

In this Letter
we present a general and highly effective formulation 
of moment equations,
expressed in terms of the {\it binomial moments},
defined below.
These are linear combinations of the ordinary moments,
that capture the structure of the reactions and provide
much physical insight. 
As a result,
this formulation enables the inclusion of
moments up to any desired order. 
The equations are linear and their number 
is dramatically reduced compared to the master equation,
enabling the analysis of complex reaction networks.

Consider a reaction network, consisting of 
$J$
reactive species,
$X_i$, $i=1,\dots,J$.
These species undergo reactions of the form

\begin{equation}
\sum_{i=1}^{J}{n_i X_i} \rightarrow \sum_{j=1}^{J}{m_j X_j},
\label{eq:chemical_reaction}
\end{equation}

\noindent
where the stoichiometric coefficients 
$n_i$ and $m_i$
are integers.
The order of the reaction is given by
$n = \sum_{i=1}^{J}{n_i}$,
namely the number of reactants on the left hand side of
Eq. (\ref{eq:chemical_reaction}).
In the context of chemical reactions it is common to omit reactions 
of order higher than 
$n = 2$.
However, in the formulation presented below there is no such
limitation.
The reactions presented in 
Eq. (\ref{eq:chemical_reaction})
can be  expressed in a vector form as
$\vec n \rightarrow \vec m$,
namely a set of molecules with stoichiometric coefficients 
$\vec n$, 
reacts to form a new set
$\vec m$.
The rate constant for the molecular configuration
$\vec n$ 
to react is given by
$\T n$ 
(s$^{-1}$).
Certain reactions may take several 
paths, with different probabilities or branching ratios.
The probability that the reaction will result in the configuration
$\vec m$ 
is denoted by
$\P nm$.
These probabilities satisfy
$\sum_{\vec m}{\P nm} = 1$.
Thus, the rate constant for the reaction 
$\vec n \rightarrow \vec m$
is
$\T n \P nm$.

In the stochastic analysis one describes the state of  
the system by the vector
$\vec N = (N_1,\dots,N_J)$,
where
$N_i$
is the number of copies of species 
$X_i$.
Below we introduce a combinatorial approach, 
which will later allow us
to express the master equation and the moment equations
in a transparent form.
Let 
$\vec v = (v_1,\dots,v_J)$
be a vector of integers.
It can be expressed as a linear 
combination of the basis vectors,
$\vec e_1,\dots,\vec e_J$,
where
$\vec e_i = (0,\dots, v_i = 1, \dots, 0)$.
We denote by 
$\Q v$
a combination of molecules consisting of exactly
$v_i$
copies of the species 
$X_i$.
The number of such combinations that exist in a system at the state
$\vec N$
is given by

\begin{equation}
\WN v = {{\vec N} \choose {\vec v}},
\label{eq:WNv}
\end{equation}

\noindent
where 
${\vec N \choose{\vec v}} = \prod_{i=1}^{J}{{N_i \choose{v_i}}}$
and 
${N_i \choose{v_i}}$
is the binomial coefficient.
To illustrate the motivation for this definition, 
consider 
the reaction 
$\vec n \rightarrow \vec m$.
It occurs at a rate proportional to 
$\T n \P nm$,
and to the number of 
$\Q n$
combinations 
which are
present in the system, 
given by $\WN n$.

Let 
$\Prob$
represent the time dependent probability for the system to be in the state
$\vec N$. 
The master equation for $\Prob$
takes the form

\begin{eqnarray}
&&\frac{d\Prob}{dt} =
\sum_{\vec n, \vec m}{
\T n \P nm 
\left[
{\bf W}(\vec N + \vec n - \vec m, \vec n)
P(\vec N + \vec n -\vec m) - 
\WN n \Prob
\right]
}.
\label{eq:master}
\end{eqnarray}

\noindent
In this equation one sums over all the reactions
$\vec n \rightarrow \vec m$
in the system.
These reactions yield a positive contribution to
$\Prob$
when the state of the system is
$\vec N + \vec n - \vec m$,
and a negative contribution to
$\Prob$ 
when the system is in the state
$\vec N$.
In numerical simulations the master equation must be truncated
in order to maintain a finite number of equations.
This is achieved by setting upper cutoffs
$C_i$,
$i = 1, \dots, J$,
such that 
$\Prob = 0$
if
$N_i > C_i$
for any value of $i$.
The number of coupled equations, 
$N_E = \prod_{i=1}^J (C_i + 1)$,
grows exponentially
with the number of reactive species,
$J$.
This severely limits the applicability of the master equation
to complex reaction networks
\cite{Stantcheva2003,Lipshtat2004}.

A more compact description of stochastic reaction networks can be
obtained using moment equations. 
These equations, derived by 
tracing over the master equation, 
consist of ordinary differential equations for the time 
derivatives of the moments
$\av {N_1^{a_1} \cdots N_J^{a_J}}$,
where
$a_i$ 
are integers.
The order of a specific moment is given by
$k = \sum_{i=1}^{J}{a_i}$.
The difficulty with moment equations 
is that higher order moments
appear on the right hand side
of each equation.
To obtain a closed set of equations one needs to apply
a suitable truncation scheme, in which the higher
order moments are expressed in terms of low order moments 
\cite{McQuarrie1967,Gomez-Uribe2007,Barzel2007a,Barzel2007b}.
In practice, the truncation is typically done at the
level of third order moments, namely only the first and second
order moments are taken into account.
Making the truncation at higher orders turns out to be
prohibitively complicated even for relatively simple networks.
Below we introduce a different formulation of the moment
equations, which enables to extend the truncation up
to any desired order.

Consider a reaction network described by the
probability distribution $\Prob$.
The binomial moment
$\W v$
is defined as the average number of combinations
of the form
$\Q v$
that appear in 
the system.
It is given by

\begin{equation}
\W v = \sum_{\vec N}{\WN v \Prob}.
\label{eq:BM}
\end{equation}

\noindent
To understand the meaning of the binomial moments, 
consider the case where
$\vec v = \vec e_i$.
Here the corresponding binomial moment,
$\W {e_i}$,
is given by the average number of combinations 
of a single copy of the species 
$X_i$.
This is simply the average population size, 
$\N i$.
In case that  
$\vec v = \vec e_i + \vec e_j$,
the corresponding moment is
$\W v = \NN ij$,
which stands for the average number of
$X_i$-$X_j$ 
pairs present in the system.
Note that for 
$\vec v = 2 \vec e_i$
the corresponding binomial moment is
$\W v = (\Ns i - \N i)/2$.
Similarly, the number of 
$X_i$ 
triplets is given by 
$\W v = (\Nc i - 3\Ns i + 2\N i)/6$,
where
$\vec v =  3 \vec e_i$.
The order the binomial moment
$\W v$ 
is
$k = \sum_{i=1}^{J} v_i$.

To obtain the binomial moment equations 
we express the time derivative of 
$\W v$ 
using Eq. (\ref{eq:BM}),
where the time derivative of 
$\Prob$
is taken from the master equation 
[Eq. (\ref{eq:master})],
and a summation is taken over
$\vec N$
\cite{LongBME}.
The resulting binomial moment
equations take the form

\begin{equation}
\frac{d\W v}{dt} =
\sum_{\vec n, \vec m}
\left[
\B nm
{{\vec v + \vec n - \vec m}\choose{\vec n}}
\Wm {\vec v + \vec n - \vec m}
- \T n 
{{\vec n}\choose{\vec m}}
{{\vec v + \vec m}\choose{\vec n}}
\Wm {\vec v + \vec m}
\right],
\label{eq:BME}
\end{equation}

\noindent
where

\begin{equation}
\B nm = \sum_{\vec w}
{\T n \P nw {\vec w \choose \vec m}}.
\label{eq:Bnm}
\end{equation}

\noindent
The first term in Eq.
(\ref{eq:BME})
accounts for positive contributions
of the reactions to 
$\W v$,
while the second term acounts for the negative contributions.
We first consider the positive contributions.
Consider a single combination 
$\Qm {\vec v + \vec n  - \vec m}$.
There are on average 
$\Wm {\vec v + \vec n  - \vec m}$
such combinations in the system.
For each one of these combinations,
the reaction 
$\vec n \rightarrow \vec m$,
produces a new 
$\Q v$ 
combination, 
so $\W v$ increases.
The rate constant 
$\B nm$ 
accounts for the rate of formation
of 
$\Q m$
combinations by the reactions
$\vec n \rightarrow \vec w$
for all possible choices of
$\vec w$.
The reaction rate is also proportional 
to the binomial coefficient
${{\vec v + \vec n - \vec m}\choose{\vec n}}$,
that accounts for the number of combinations
$\Q n$ 
in
$\Qm {\vec v + \vec n  - \vec m}$,
each of which may undergo the reaction.

We now refer to the negative contributions.
Consider the combination 
$\Qm {\vec v + \vec m}$
which undergoes a reaction of the form
$\vec n \rightarrow \vec w$,
for any possible choice of $\vec w$.
The overall rate of these reactions
is given by
$\T n {{\vec v + \vec m}\choose{\vec n}}$.
Each time such a reaction takes place,
a single combination
$\Q n$ 
is removed from the system.
The removed combination
$\Q n$ 
can be decomposed into
$\Q m$
and
$\Qm {\vec n - \vec m}$.
Note that there are 
${{\vec n}\choose{\vec m}}$
different possibilities to perform this decomposition of
$\Q n$.
When 
$\Q m$
is removed,
the combination 
$\Qm {\vec v + \vec m}$
is replaced by
by 
$\Q v$.
This 
$\Q v$
combination is then eliminated by the
subsequent removal of
$\Qm {\vec n - \vec m}$.

Eq. (\ref{eq:BME}) is not in a closed form, because higher order
moments appear on the right hand side of each equation.
However, the binomial moments tend to decrease as their
order increases. 
To demonstrate this feature, consider a binomial moment
$\W v$ 
of order
$k = \sum_{i=1}^J v_i$.
It represents the average number of appearances of a certain combination
$\Q v$
which consists of 
$k$
molecules in the system.
In a small system, 
where the average copy numbers are small,
$\W v$ 
tends to decrease as $k$ increases.
This enables us to use the following truncation scheme.
We choose a cutoff $C$ such that
$\W v$
is set to zero whenever
$k > C$.
The number of different binomial moments of order
$k$
is given by
${{k + J - 1}\choose{k}}$.
Thus the number of binomial moment equations, 
after the truncation is carried out, is 
$N_E = \sum_{k=1}^{C}{{k + J - 1}\choose{k}}$.
While the number of equations in the 
master equation grows exponentially with $J$,
the number of binomial moment equations
scales only polynomially with $J$.
Moreover, in practice one can obtain accurate results
using surprisingly low values of the cutoffs.
In fact, in an earlier version of this formulation,
it was shown that for a broad range of conditions 
the cutoff
$C = 2$
is sufficient
\cite{Barzel2007a,Barzel2007b}.
In this case, the equations include first order moments that
account for the average population sizes and second order moments
that account for the number of pairs of species $X_i$ and $X_j$
in the system, from which the reaction rates are evaluated.
In cases where a cutoff of 
$C = 2$
is not sufficient, 
one may raise the cutoff
until accurate results are obtained.
In the small system limit the required cutoff is usually low.
In the large system limit, stochastic equations are no longer
required and can be replaced by the rate equations.
 
\begin{figure}
\includegraphics[width=9cm]{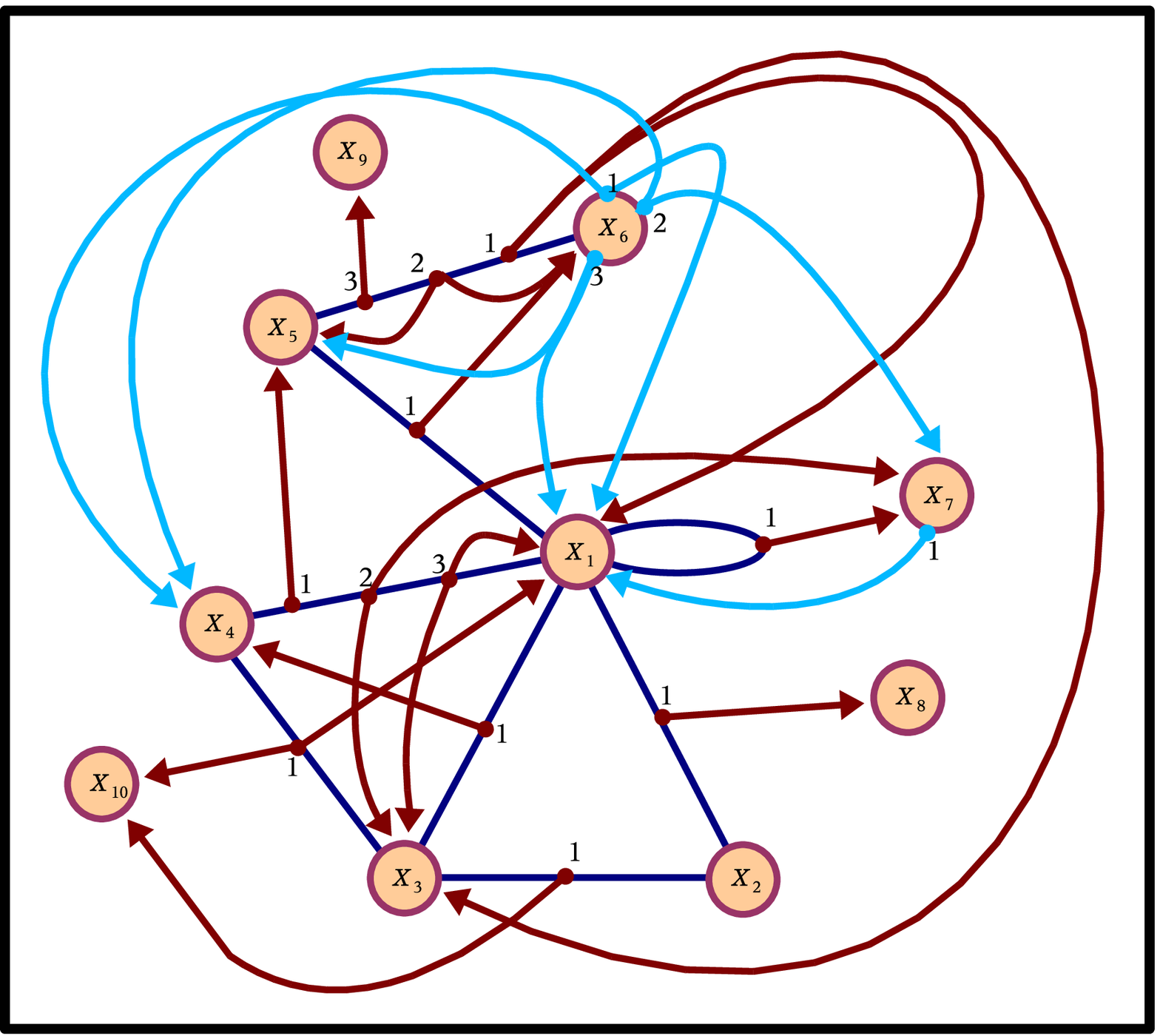}
\caption{
(Color online) 
A reaction network consisting of 
$10$
reactive species.
First order reactions 
appear as arrows connecting the reacting species to its products.
Second order reactions are denoted by 
edges connecting the two reacting species.
Arrows are drawn from the edge to the reaction products.
Different reaction paths are marked by the indices appearing by the arrows.
}
\label{fig1}
\end{figure}

To demonstrate the method, we apply it 
to the reaction network shown in
Fig. \ref{fig1}. 
This network consists of processes involving single molecules as well
as pairs of molecules. 
The network includes 
$10$
reactive species,
$3$
zero order reactions,
$14$
first order reactions and 
$12$
second order reactions.
The zero order reactions lead to the formation of
$X_1$, 
$X_2$ 
and 
$X_3$
molecules,
where
$P_0^{\vec e_i} = 1/3$
for $i=1,2$ and 3. 
The rest of the species are formed via first and second order reactions.
The first order reactions include the degradation of each of the 
reactive species, and the dissociation of 
$X_6$
and 
$X_7$.
The four first order reactions involving 
$X_6$
are given by
$\vec e_6 \rightarrow 0$ (degradation),
$\vec e_6 \rightarrow 2 \vec e_1 + \vec e_4$,
$\vec e_6 \rightarrow \vec e_4 + \vec e_7$
and
$\vec e_6 \rightarrow \vec e_1 + \vec e_5$,
where
$P_{\vec e_6}^{0} = 0.999$
and
$P_{\vec e_6}^{2 \vec e_1 + \vec e_4} = 
P_{\vec e_6}^{\vec e_4 + \vec e_7} =
P_{\vec e_6}^{\vec e_1 + \vec e_5}/2 = 0.00025$.
The two first order reactions involving 
$X_7$
are 
$\vec e_7 \rightarrow 0$
and
$\vec e_7 \rightarrow 2 \vec e_1$,
where
$P_{\vec e_7}^{0} = 0.999$,
$P_{\vec e_7}^{2 \vec e_1} = 0.001$.
The second order reactions involve all the pairs of nodes connected by edges.
The reaction of
$X_1$ 
and
$X_4$
includes three reaction paths,
$\vec e_1 + \vec e_4 \rightarrow \vec e_5$,
$\vec e_1 + \vec e_4 \rightarrow \vec e_3 + \vec e_7$
and
$\vec e_1 + \vec e_4 \rightarrow 2 \vec e_1 + \vec e_3$,
where
$P_{\vec e_1 + \vec e_4}^{\vec e_5} = 0.25$,
$P_{\vec e_1 + \vec e_4}^{\vec e_3 + \vec e_7} = 0.5$
and
$P_{\vec e_1 + \vec e_4}^{2 \vec e_1 + \vec e_3} = 0.25$.
The paths for the reaction of
$X_5$ 
and
$X_6$
are 
$\vec e_5 + \vec e_6 \rightarrow 5 \vec e_1 + 2 \vec e_3$,
$\vec e_5 + \vec e_6 \rightarrow \vec e_5 + \vec e_6$
and
$\vec e_5 + \vec e_6 \rightarrow \vec e_9$,
with the probabilities
$1/4$,
$1/4$
and
$1/2$,
respectively.
This means that
upon encounter of a pair of $X_5$ and $X_6$ molecules,
they either 
dissociate into their fundamental components, 
remain unchanged
or combine to form the molecule
$X_9$.
To characterize the size of the system, 
we introduce the parameter
$S$.
The rate of zero order reactions is taken to be 
proportional to the size of the system, 
while the rate of
second order reactions is inversely proportional to the size of the system
\cite{Lohmar2009}.
We thus set
$T_0 = g S$,
$\T {e_i} = d_i$
and
$T_{\vec e_i + \vec e_j} = a_{ij}/S$.
The parameters we use are
$g = 1$ s$^{-1}$
for the zero order reactions, 
$d_i = 0.3$ s$^{-1}$,
$i=1,\dots,3$,
for the first order reactions and
$a_{ij} = 1$ s$^{-1}$
for the second order reactions
that are included in the network.
 
For this reaction network, setting a cutoff of 
$C = 3$
for all the species, one obtains 
$(3 + 1)^{10} \approx 10^6$ 
equations in the master equation.
In contrast, 
the same cutoff set in the binomial moment equations yields only
$285$ 
equations. 
A lower cutoff of
$C = 2$
results in
$65$
equations, 
compared with approximately 
$6 \times 10^4$
equations in the master equation.
After solving the binomial moment equations, 
one can extract the average population sizes,
which are given by the
$\W {e_i}$ 
moments.

\begin{figure}
\includegraphics[width=8cm]{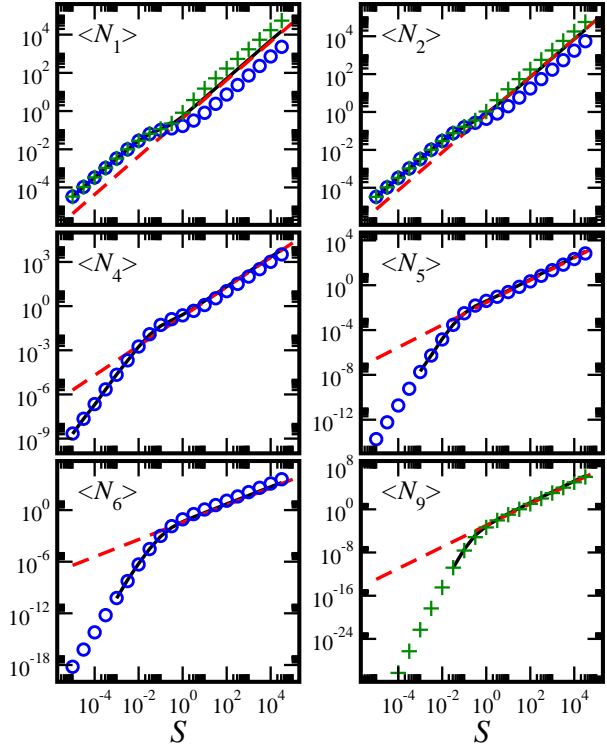}
\caption{
(Color online)
The average populations sizes, 
$\N i$, 
of several species from the reaction network appearing in
Fig. \ref{fig1},
vs. the system size, 
$S$,
as obtained from the moment equations under steady 
state conditions (circles).
These results were obtained for a cutoff of 
$C = 2$.
In the limit of small system sizes the results 
are in perfect agreement with the
results obtained from the master equation (solid lines).
The rate equation results (dashed lines) show 
significant deviations for small sizes.
For the species 
$X_1$
and
$X_2$
the moment equations deviate for large systems,
and in order to get accurate results one has to use a higher cutoff.
For these species we display results obtained 
from the moment equations with a cutoff of
$C = 3$ ($+$). 
The species 
$X_9$
cannot be produced without the presence of triplets in the system
(bottom right display).
A cutoff of 
$2$ 
is thus insufficient for obtaining its population size.
However using the moment equations with a cutoff of 
$3$ ($+$)
provides accurate results for the entire range of system sizes. 
}
\label{fig2}
\end{figure}

\begin{figure}
\includegraphics[width=6.5cm]{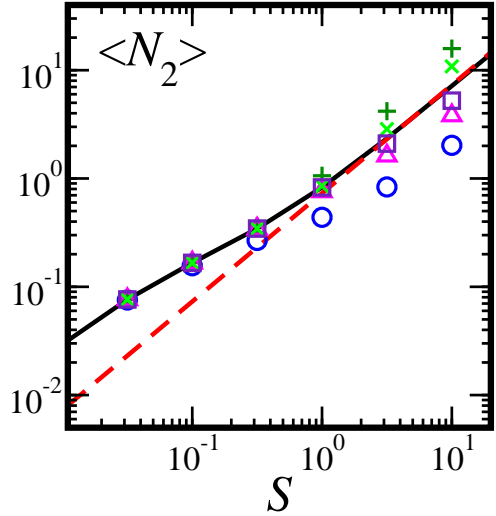}
\caption{
(Color online)
Here we focus on the transition between the stochastic regime
$S < 1$
and the deterministic regime 
$S > 1$.
Results are shown for
$\N 2$
obtained from the moment equations with a cutoff of
$2$ (circles),
$3$ ($+$),
$4$ (triangles),
$5$ ($\times$)
and
$6$ (squares).
As the cutoff used in the moment equations is raised, 
the convergence to the rate equation results 
(dashed line) becomes smoother.
When a cutoff higher than 
$6$
is required,
one reliably enters the range of validity of the rate equations.  
}
\label{fig3}
\end{figure}

In Fig. \ref{fig2}
we show the population sizes of several selected species 
vs. the system size,
$S$.
These results were obtained
from the binomial moment equations
with $C = 2$ (circles),
under steady state conditions.
For small systems, 
the results are in perfect agreement with those of the
master equation (solid lines).
The master equation results were obtained 
using the Gillespie algorithm
\cite{Gillespie1977}, since 
direct integration of the master equation is already
infeasible in this case
\cite{Rareevents}.
For large systems, the stochastic results converge 
to those obtained from the rate equations (dashed lines).
For most species, 
the results of the binomial moment equations with such low cutoff, 
coincide with those of the master equation
not only in the small system limit but also
for $S \gg 1$.
In fact, for certain species, it is sufficient to choose a cutoff of 
$C = 2$
to account for the abundances in the entire range of system sizes.
This surprising feature was discussed in an earlier formulation of
the moment equations,
presented in Refs. 
\cite{Barzel2007a,Barzel2007b}.
In case that the the cutoff of
$C=2$
in insufficient one may raise it to 
$3$
or 
$4$,
until a smooth convergence to the rate equation results is obtained
in the large system limit.
Results obtained for a cutoff of
$C=3$ ($+$)
are shown for
$\N 1$
and
$\N 2$.
Consider the population of the species 
$X_9$ in the system.
This species is produced by the reaction
$\vec e_5 + \vec e_6 \rightarrow \vec e_9$.
Thus, 
in order for $X_9$ to be produced, 
a pair of 
$X_5$
and
$X_6$ 
must be simultaneously present in the system.
However, such pairs form only when there are triplets in the system.
For instance, when the combination
$\Q {e_1 + \vec e_4 + \vec e_6}$
transforms into 
$\Q {e_5 + \vec e_6}$
through the second order reaction 
$\vec e_1 + \vec e_4 \rightarrow \vec e_5$.
Thus, a cutoff lower than 
$C = 3$ 
in the binomial moment equations would terminate the production of 
$X_9$ 
even in the stochastic limit.
In such cases, 
the ability to extend the equations to higher cutoffs is crucial.
In Fig. \ref{fig3}
we show the convergence of the binomial moment equations
to the rate equation results (dashed lines) 
for the average population of $X_2$.
Results are shown for cutoffs of
$C = 2$ (circles),
$C = 3$ ($+$),
$C = 4$ (triangles),
$C = 5$ ($\times$)
and
$C = 6$ (squares).
Already for 
$C = 5$
or
$6$
the convergence to the deterministic results is smooth.
When a higher cutoff is required, one can safely use the rate equations.

In conclusion, the binomial moment equations provide a highly efficient
equation-based methodology for the simulation of stochastic reaction 
networks.
The binomial moments capture the essence of the combinatorics
that governs the reaction rates, reflecting the stoichiometric
structure of the reactions.
The closure scheme is fully controlled and determined by the maximal
number of particles allowed to reside simultaneously in the system,
which is clearly limited by the system size.
Unlike the ordinary moment equations, the binomial moment equations
can be constructed to any order using an automated procedure,
taking the network structure as input.
The number of equations is dramatically reduced. 
This method opens the way to systematic studies of 
large and complex stochastic networks 
beyond the feasibility limit of existing methods.
Moreover,
as an equation-based paradigm, it
is amenable to analytical treatments 
that are expected to provide crucial insight 
about the networks.

Baruch Barzel thanks the
Hoffman Leadership and Responsibility program 
at the Hebrew University for support.
This work was supported by the US-Israel Binational Science
Foundation.

\clearpage
\newpage

\clearpage
\newpage

\begin{thebibliography}{10}

\bibitem{Charnley1998}
{S.B. Charnley}, 
Astrophys. J. {\bf 509},  {L121}  (1998).

\bibitem{McAdams1997}
{H.H. McAdams and A. Arkin}, 
Proc. Natl. Acad. Sci. US {\bf 94},  814  (1997).

\bibitem{Kaern2005}
{M. Kaern, T.C. Elston, W.J. Blake and J.J. Collins}, 
Nature Review Genetics
{\bf 6},  451  (2005).

\bibitem{Lande2003}
R. Lande, S. Engen, B.E. S{\ae}ther,
{\em Stochastic Population Dynamics in Ecology and Conservation}
(Oxford University Press, New York, 2003).

\bibitem{vanKampen1981}
{N.G. van Kampen}, 
{\em {Stochastic Processes in Physics and Chemistry}}
({North-Holland}, {Amsterdam}, 1981).

\bibitem{Gardiner1985}
{C.W. Gardiner}, 
{\em {Handbook of Stochastic Methods}} 
({Springer-Verlag}, {Berlin}, 2004).

\bibitem{Gillespie1977}
{D.T. Gillespie}, 
{J. Phys. Chem.} {\bf 81}, 2340 (1977).

\bibitem{Gillespie2007}
{D.T. Gillespie}, 
{Annu. Rev. Phys. Chem.} 
{\bf 58},  35  (2007).

\bibitem{Stantcheva2003}
{T. Stantcheva and E. Herbst}, 
Mon. Not. R. Astron. Soc. 
{\bf 340},  983  (2003).

\bibitem{Lipshtat2004}
{A. Lipshtat, O. Biham}, 
Phys. Rev. Lett. {\bf 93}, 170601  (2004).

\bibitem{McQuarrie1967}
{D.A. McQuarrie}, 
{J. Appl. Prob.} {\bf 4},  413  (1967).

\bibitem{Gomez-Uribe2007}
{C.A. G\'omez-Uribe and G.C. Verghese}, 
J. Chem. Phys. {\bf 126},  024109   (2007).

\bibitem{Barzel2007a}
{B. Barzel and O. Biham}, 
Astrophys. J. {\bf L115},  20941  (2007).

\bibitem{Barzel2007b}
{B. Barzel and O. Biham}, 
J. Chem. Phys. {\bf 127},  144703  (2007).

\bibitem{LongBME}
{The details of this derivation will appear in B. Barzel and O. Biham,
to be submitted to Phys. Rev. E}.

\bibitem{Lohmar2009}
{I. Lohmar, J. Krug and O. Biham}, 
Astron. Astrophys. {\bf 504},  {L5}  (2009).

\bibitem{Rareevents}
{Note that for some species the solid line in Fig. \ref{fig2} does not
extend all the way to the left. This is due to the fact that the relevant
reactions become rare in the small system limit and cannot be evaluated by
the Gillespie algorithm.}

\end{thebibliography}
\end{document}